\documentclass[acmlarge]{acmart}

\AtBeginDocument{%
  \providecommand\BibTeX{{%
    \normalfont B\kern-0.5em{\scshape i\kern-0.25em b}\kern-0.8em\TeX}}}

\setcopyright{acmcopyright}
\copyrightyear{2020}
\acmYear{2020}

\usepackage{caption}
\usepackage{subfigure}
\usepackage{amsmath}
\usepackage{array}
\usepackage{url}

\begin{document}

\title{$C^3DRec$: Cloud-Client Cooperative Deep Learning for Temporal Recommendation in the Post-GDPR Era}

\author{Jialiang Han}
\affiliation{%
  \institution{Peking University}
  \streetaddress{5 Yiheyuan Rd}
  \city{Beijing}
  \country{China}
}
\email{hanjialiang@pku.edu.cn}

\author{Yun Ma}
\affiliation{%
  \institution{Peking University}
  \streetaddress{5 Yiheyuan Rd}
  \city{Beijing}
  \country{China}
}
\email{mayun@pku.edu.cn}

\renewcommand{\shortauthors}{Han, et al.}

\begin{abstract}
  Mobile devices enable users to retrieve information at any time and any place. Considering the occasional requirements and fragmentation usage pattern of mobile users, temporal recommendation techniques are proposed to improve the efficiency of information retrieval on mobile devices by means of accurately recommending items via learning temporal interests with short-term user interaction behaviors. However, the enforcement of privacy-preserving laws and regulations, such as the General Data Protection Regulation (GDPR), may overshadow the successful practice of temporal recommendation. The reason is that state-of-the-art recommendation systems require to gather and process the user data in centralized servers but the interaction behaviors data used for temporal recommendation are usually non-transactional data that are not allowed to gather without the explicit permission of users according to GDPR. As a result, if users do not permit services to gather their interaction behaviors data, the temporal recommendation fails to work. To realize the temporal recommendation in the post-GDPR era, this paper proposes $C^3DRec$, a cloud-client cooperative deep learning framework of mining interaction behaviors for recommendation while preserving user privacy. $C^3DRec$ constructs a global recommendation model on centralized servers using data collected \textit{before} GDPR and fine-tunes the model directly on individual local devices using data collected \textit{after} GDPR. We design two modes to accomplish the recommendation, i.e., \textit{pull mode} where candidate items are pulled down onto the devices and fed into the local model to get recommended items, and \textit{push mode} where the output of the local model is pushed onto the server and combined with candidate items to get recommended ones. Evaluation results show that $C^3DRec$ achieves comparable recommendation accuracy to the centralized approaches, with minimal privacy concern.
\end{abstract}

\ccsdesc[500]{Information systems~Recommender systems}
\ccsdesc[300]{Computing methodologies~Machine learning}
\ccsdesc[500]{Human-centered computing~Mobile devices}

\keywords{Temporal Recommendation, Cloud-Client Cooperative Training, GDPR}

\maketitle

\section{Introduction}
Due to the information explosion, recommending interesting information to users becomes more and more important~\cite{shapira2015recommender}. With the rapid growth and widespread of smartphones and tablet computers, users are able to access the Internet to retrieve various kinds of information and dive into oceans of applications without the restriction of time and locations~\cite{li2020systematic, xu2020approximate, chen2018through}. The special characteristics of user interactions on mobile devices call for new requirements of recommendation services. On one hand, users are likely to raise occasional requirements according to the situation they stay. As a result, recommendation services should efficiently capture users’ emerging interests to meet such a kind of temporal information demand. On the other hand, mobile usage is usually fragmented\cite{karlson2010mobile, liu2017understanding, li2015characterizing}, meaning that users interact with mobile devices in a short period. As a result, the recommendation should be achieved by a very small amount of data.

To improve the accuracy of recommendation services on mobile devices, temporal recommendation techniques~\cite{song2016multi, tu2019fingerprint, imai2018early} have been proposed by leveraging temporal user behavior data generated on mobile devices~\cite{wang2019modeling, lu2018inferring, CHENDLDEPLOY2}, such as item clicks, dwell time, and revisitation frequency. 

As far as we are concerned, most of the existing approaches \cite{hidasi2015session, hidasi2016parallel, wu2017recurrent} and industrial practices \cite{liu2012enlister, davidson2010youtube} require to upload user behavior data from individual devices to the cloud server of recommendation service, where a recommendation model can be trained. Then, the recommendation process is performed either on the cloud or individual devices if the trained model is downloaded.

However, the above practices raise privacy issues. There are a considerable number of discussions about the possession, usage, and portability of user-generated data. Those public concerns about privacy leakage have gradually influenced policy and legislation formulation and eventually lead to the General Data Protection Regulation (GDPR)\footnote{https://gdpr-info.eu/}, California Consumer Privacy Act (CCPA)\footnote{https://oag.ca.gov/privacy/ccpa/} and other privacy laws and regulations. GDPR states that a specific, freely-given, plainly-worded, and unambiguous consent should be given by the data subject, i.e. the user, and the data subject has the right of access, the right to data portability and the right to be forgotten. As a result, in the post-GDPR era, if the user refuses to upload their behavior data to train a global recommendation model, the above approaches would be impractical.

Since the user behavior data produced \textit{after} GDPR cannot be uploaded without the user's consent, a straight forward solution is to take advantage of prior knowledge. DeepType \cite{xu2018deeptype} trains a global model on the cloud using massive public corpora, and then incrementally customizes the global model with data on individual devices, i.e. trains a personalized model for each user on his device. However, due to the heterogeneous of item metadata, i.e. item ID, category ID, there is few public dataset of cross-platform recommendation, to extract common prior knowledge for most recommendation systems, which means that we have to rely on existing behavior data collected \textit{before} GDPR to extract common knowledge. This raises the question of whether it complies with GDPR to process existing data collected \textit{before} GDPR. After carefully going through GDPR, we surprisingly find out that there is an exemption called Legitimate Interests\footnote{https://ico.org.uk/for-organisations/guide-to-data-protection/guide-to-the-general-data-protection-regulation-gdpr/lawful-basis-for-processing/legitimate-interests/} in GDPR which allows processing existing user data for legitimate interests without users' consent under careful consideration. The application scene of this mechanism includes fraud detection, crime prediction, network security protection, and marketing guidance, which means that the Legitimate Interests mechanism allows recommendation service to process user behavior data collected \textit{before} GDPR.

Inspired by related work~\cite{huang2016programming} and Legitimate Interests exemption of GDPR, we propose $C^3DRec$, a cloud-client cooperative deep learning framework for temporal recommendation in the post-GDPR era. First, we train a Recurrent Neural Network (RNN) based global recommendation model with existing user behavior data collected \textit{before} GDPR for all users on the cloud, and push the global model to individual devices. Second, we train an RNN-based personalized recommendation model with real-time user behavior data collected \textit{after} GDPR for each user on his device, and extract a unique user embedding from his personalized model. Third, individual devices pull a recommendation item candidate set from the cloud (\textbf{pull mode}) or push user embeddings toward the cloud (\textbf{push mode}), and then we combine the latent information of user embeddings and item embeddings to complete the recommendation process. However, it is not straight-forward to deploy deep learning based applications on mobile devices~\cite{chen2020comprehensive, xu2019first}. Specifically, in this cloud-client cooperative recommendation scenario, there remain two challenges to tackle. To reduce network communication overhead between the cloud and devices, we perform the Least Absolute Shrinkage and Selection Operator (Lasso) regression, or $L_1$ regularization, to sparsify the item embeddings of item candidate set in pull mode or user embeddings in push mode. To reduce computational overhead on the devices, we perform an Automated Gradual Pruner (AGP)\cite{zhu2017prune} to compress the recommendation model.

To evaluate the performance of the proposed framework, we conduct experiments in terms of accuracy, communication overhead and computational overhead, based on a public user behavior dataset from Taobao\cite{zhu2018learning}, a large-scale e-commercial application in China, which contains nearly a million users and nearly 100 million interactions of users and items. The experimental results show that we achieve up to 10x reduction in communication overhead, reduce the computational overhead of on-device training towards the range of computational resources of middle-class mobile devices, with minimal loss in accuracy. We summary our contributions of $C^3DRec$ as follows.
\begin{itemize}
\item We propose a cloud-client cooperative deep learning framework for temporal recommendation, which alleviates the risks of user privacy leakage and violation of privacy-related laws, in benefit of recommendation service providers.
\item We reduce both communication and computational overhead of individual devices by AGP weight pruning and Lasso regression ($L_1$ regularization), to make sure that it is more feasible and less resource-consuming to perform on-device training of personalized models on real-world mobile devices, in benefit of application developers.
\item We evaluate the performance of $C^3DRec$ on a large-scale public dataset, and the experimental results show that we achieve up to 10x reduction in communication overhead and reduce the computational overhead towards the range of middle-class mobile devices, with minimal loss in accuracy.
\end{itemize}

The rest of the paper is organized as follows. In Section 2, we formally define the problem of temporal recommendation in the post-GDPR era. In Section 3, we present the general framework of $C^3DRec$ and optimizations to reduce communication and computational overhead. In Section 4, we evaluate $C^3DRec$ on a large-scale e-commerce dataset to demonstrate its effectiveness and efficiency. In Section 5, we discuss several limitations of our framework and possible solutions. In Section 6, we compare $C^3DRec$ with related work. Finally, Section 7 concludes the paper.

\section{Problem Definition}
In this section, we present the problem definition of temporal recommendation in the post-GDPR era. The problem statement of recommendation is that, given a recommendation context $C$, recommendation is to learn a function $f$ which maps $C$ to the recommendation target $t: t\Leftarrow f(C)$. The recommendation context $C$ can be historical user interaction with the data source, or the data from other users, etc. The recommendation target $t$ is to predict the likelihood $p_{i,j}$ that a user $i$ would prefer an item $j$. Now let us define some important concepts in our setting:

\textbf{Definition 1 (Transactional Data):} Transactional data are user interaction data that require logging-in to produce and collect in the recommendation system, including purchasing, adding to chart, and adding to favorites, etc. We use $T_{i, t}$ to represent the item ID that user $i$ purchases, adds to chart, or adds to favorite at timestamp $t$.

\textbf{Definition 2 (Non-transactional Data):} Non-transactional data are user interaction data that do not require logging-in to produce and collect in the recommendation system, such as clicking. We use $NT_{i, t}$ to represent the item ID that user $i$ clicks at timestamp $t$.

\textbf{Definition 3 (Privacy-preserving Recommendation):} A traditional recommendation system requires uploading all interaction data to complete the recommendation, which faces the problem of privacy leakage. However, the goal of a privacy-preserving recommendation is to eliminate the uploading of any unnecessary user privacy data, i.e. non-transactional data, while still providing a customized recommendation, i.e. the next click. In order to build a privacy-preserving recommendation system, non-transactional data should be kept on-device during the whole process of recommendation, which means it is not allowed to upload any raw data or intermediate results of non-transactional data. However, the cloud can still leverage the transactional data to provide a rough recommendation candidate set.

\textbf{Definition 4 (Recommendation Item Candidate Set):} As mentioned above, a candidate set $Candidate_{i,t}$ of user $i$ at timestamp $t$, can be built or learned from all users' transactional data before timestamp $t$.

\textbf{Definition 5 (Cloud-Client Cooperative Recommendation):} Because non-transactional data are not allowed to upload during privacy-preserving recommendation, while transactional data can only help build a rough recommendation candidate set, a cloud-client cooperative recommendation system is required to learn from the local non-transactional data on the device of each user, and complete the process of personalized recommendation. Given all local non-transational data $S_{NT,i,t} = \{NT_{i, t_k}|t_k<t\}$ of user $i$ before timestamp $t$, and the candidate set $Candidate_{i,t}$ of user $i$ at timestamp $t$, the likelihood $p_{i,j,t}$ that user $i$ would click item $j$ at timestamp $t$ is predicted by the on-device recommendation model $M_{i,t}$ of user $i$ at timestamp $t$:
\begin{equation}
    p_{i,j,t}=M_{i,t}(S_{NT,i,t},Candidate_{i,t};j).
\end{equation}
Note that the on-device recommendation model $M_{i,t}$ is updated (fine tuned) from time to time for each user $i$, as the computation resources of user devices are not sufficient as those of cloud, the computational overhead of it should be relatively small to match the limited resources of devices.

If we assume that there exists a loss function $\mathcal{L}: [0,1]\times\{0,1\}\times\Theta \rightarrow \mathbb{R}$, where $\Theta$ is the parameter set domain of the on-device recommendation model, the objective of cloud-client cooperative recommendation is to solve the following optimization problem:
\begin{equation}
\begin{aligned}
&\theta^* = argmin_{\theta}\ \sum_t\sum_i\sum_j\mathcal{L}(p_{i,j,t}, label_{i,j,t}; \theta),\\
&subject\ to\ constraints\ on\ computational\ overhead,\\
\end{aligned}
\end{equation}
where $\mathcal{L}$ can be a classification loss like the cross-entropy loss, or a learning-to-rank loss\cite{rendle2012bpr, shi2012climf, steck2015gaussian}, $p_{i,j,t}$ is the predicted likelihood of user $i$ click item $j$ at timestamp $t$ by the on-device recommendation model, $label_{i,j,t} \in \{0,1\}$ is the ground truth whether item $j$ is clicked by user $i$ at timestamp $t$, $\theta$ is the parameter set of the on-device recommendation model.

We partition the whole procedure into 3 stages: global training, personalized training, testing, by $T_{device}$ and $T_{test}$. $T_{device}$ is the timestamp after which non-transactional data are used for training a personalized model for each user, instead of training a global model for all users, and $T_{test}$ is the timestamp after which labeled non-transactional data are used for testing, instead of training.

Transactional data can always be collected, whether during training or testing, therefore, the recommendation candidate set is kept up-to-date for all time, whether during training or during testing.

Before $T_{device}$, we use the non-transactional data of all users to train a global recommendation model. After $T_{device}$, we use each user's private non-transactional data to train his/her personalized recommendation model.

Before $T_{test}$, we use the non-transactional data for global training and personalized training. After $T_{test}$, we only use non-transactional data for testing.

\section{Approach}
In this section, we first introduce the working flow of both pull mode and push mode of $C^3DRec$, after that, we explain two fundamental structures in this framework in detail. In addition, we explain the optimization methods to tackle with the communication and computational overhead challenges.

\subsection{Working Flow Overview}
\begin{figure}[htbp]
  \centering
  \includegraphics[width=\textwidth]{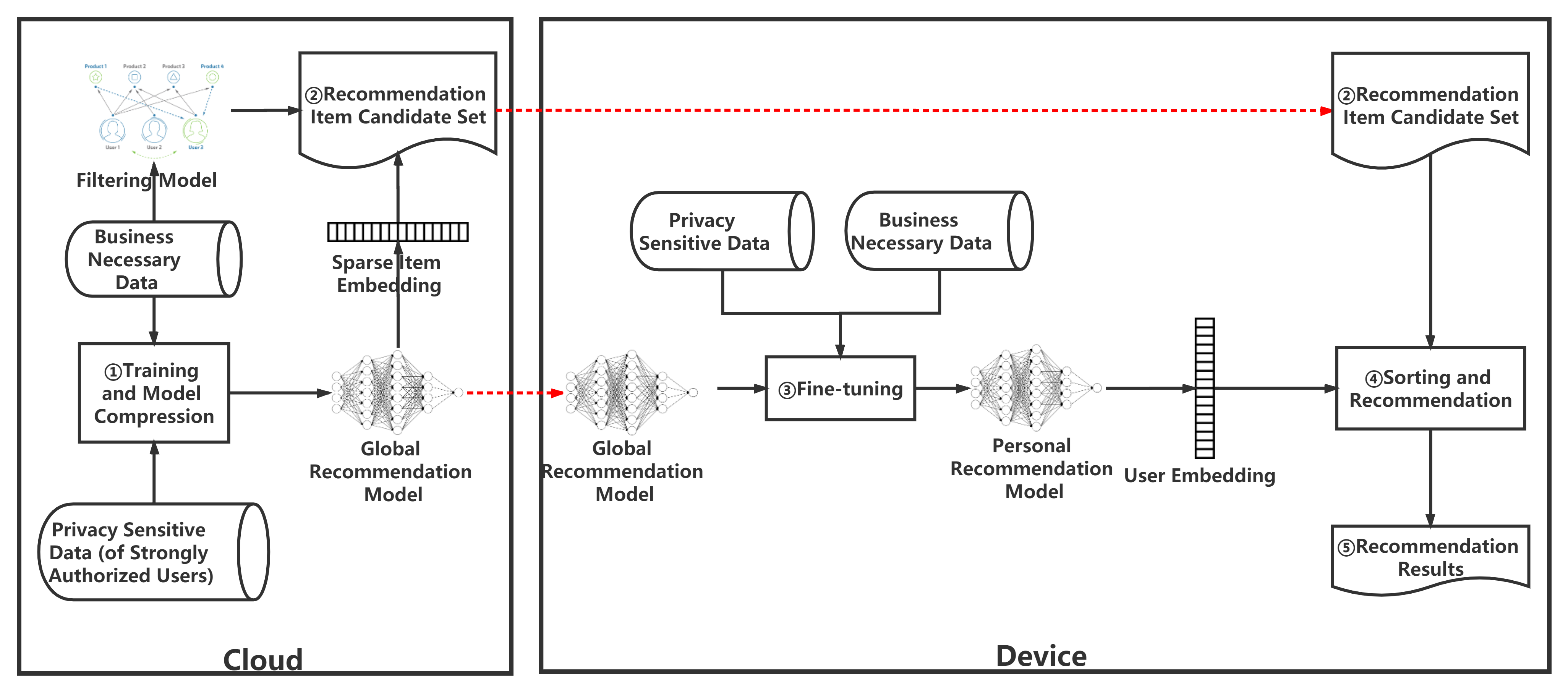}
  \caption{Pull Mode of $C^3DRec$}
  \label{design pull}
  \Description{}
\end{figure}
\begin{figure}[htbp]
  \centering
  \includegraphics[width=\textwidth]{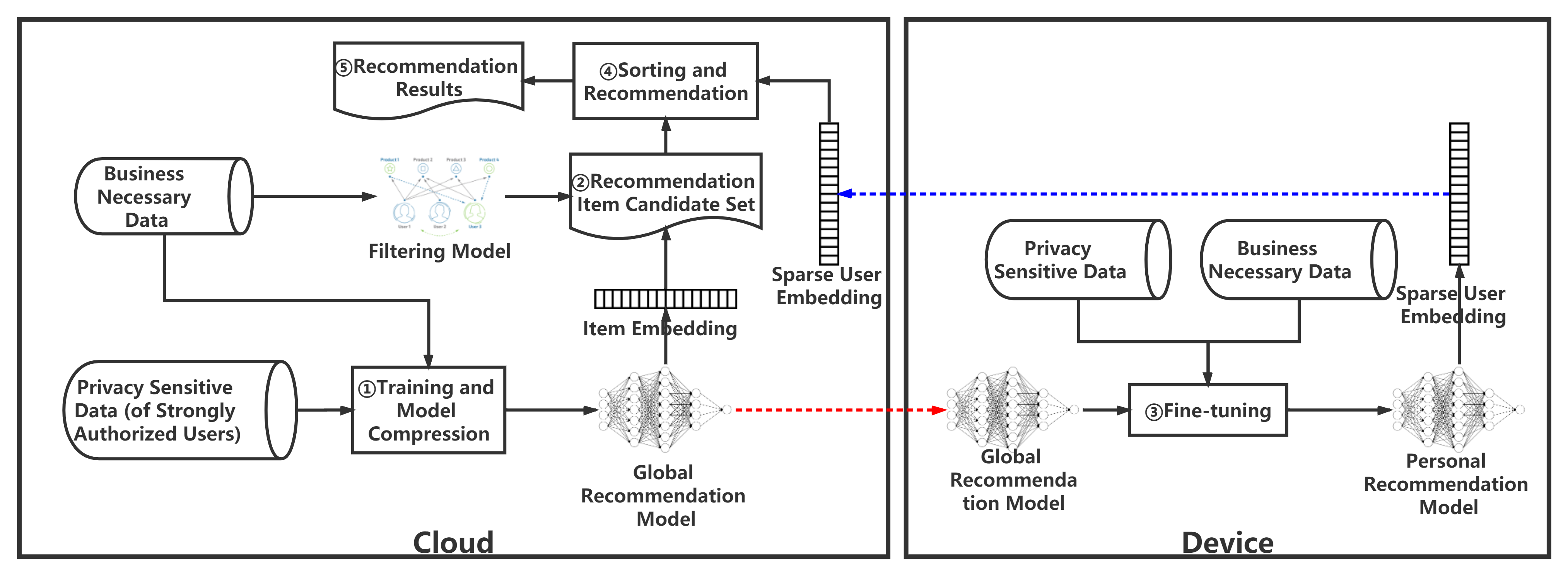}
  \caption{Push Mode of $C^3DRec$}
  \Description{}
  \label{design push}
\end{figure}
Figure \ref{design pull} and Figure \ref{design push} show the working flows of pull mode and push mode, respectively. First, we train a Gate Recurrent Unit (GRU)\cite{cho2014properties} based recommendation model, i.e. global recommendation model, with existing user behavior data collected \textit{before} GDPR for all users, and extract a recommendation item candidate set through a collaborative filtering based model on the cloud, and push the global recommendation model to individual devices. Second, we train a GRU based recommendation model, i.e. personalized recommendation model, with real-time user behavior data collected \textit{after} GDPR for each user on his device, and extract a unique user embedding from his personalized recommendation model. Third, individual devices pull a recommendation item candiate set from the cloud (in pull mode), or push user embeddings toward the cloud (in push mode), and in both modes, we combine the latent information of user embeddings and item embeddings to calculate a score representing the probability a user would visit an item in future.

\subsection{Filtering Model}
In order to reduce the computational overhead of ranking the combination of a given user embedding and recommendation item embeddings, we use a k-nearest neighbor item-based collaborative filtering (item-CF) model\cite{sarwar2001item, davidson2010youtube} to extract a recommendation item candidate set from the whole item set, whose main idea is that a user is likely to purchase an item that shares similar features, i.e. user-item interactions, with items he purchases before. The filtering model predicts a rough probability $p_{k,m}$, representing the probability user $k$ \textit{clicks} item $m$, from a user-item interactive matrix $(x_{km})_{K*M}$ where $x_{km}$ represents whether user $k$ \textit{purchases} item $m$. In particular, $p_{k,m}=\frac{\sum_{i_b}{sim(i_m,i_b)x_{k,b}}}{\sum_{i_b}\left|sim(i_m,i_b)\right|}$, where $i_m$ represents the $m^{th}$ column of the interactive matrix $(x_{km})^{K*M}$, which is a vector with a length of $K$, $sim(i_m,i_b)=\frac{i_m \cdot i_b}{\left\|i_m\right\|\left\|i_b\right\|}$.

Note that the interactive matrix is composed of only purchase records, instead of click records, which is necessary for the basic business of any online organization providing an e-commercial or likewise service. Therefore, to use at least the basic service provided like online shopping or rating movies, users have no choice but to allow uploading these purchase records, which should not be considered as privacy from a GDPR perspective. However, it is still not common practice to use an interactive matrix of \textit{purchase} to predict the action of \textit{click}. Therefore, we only use this filtering model for guidelines and obtain a candidate set from the original recommendation item set. We manually set a threshold probability $p_{candidate}$ and obtain a personalized candidate set $Candidate_k=\left\{m|p_{k,m}>p_{candidate} \right\}$.

\subsection{Recommendation Model}

\begin{figure}[htbp]
\centering
\subfigure[]{
    \begin{minipage}[t]{0.45\textwidth}
        \centering
        \includegraphics[width=\textwidth]{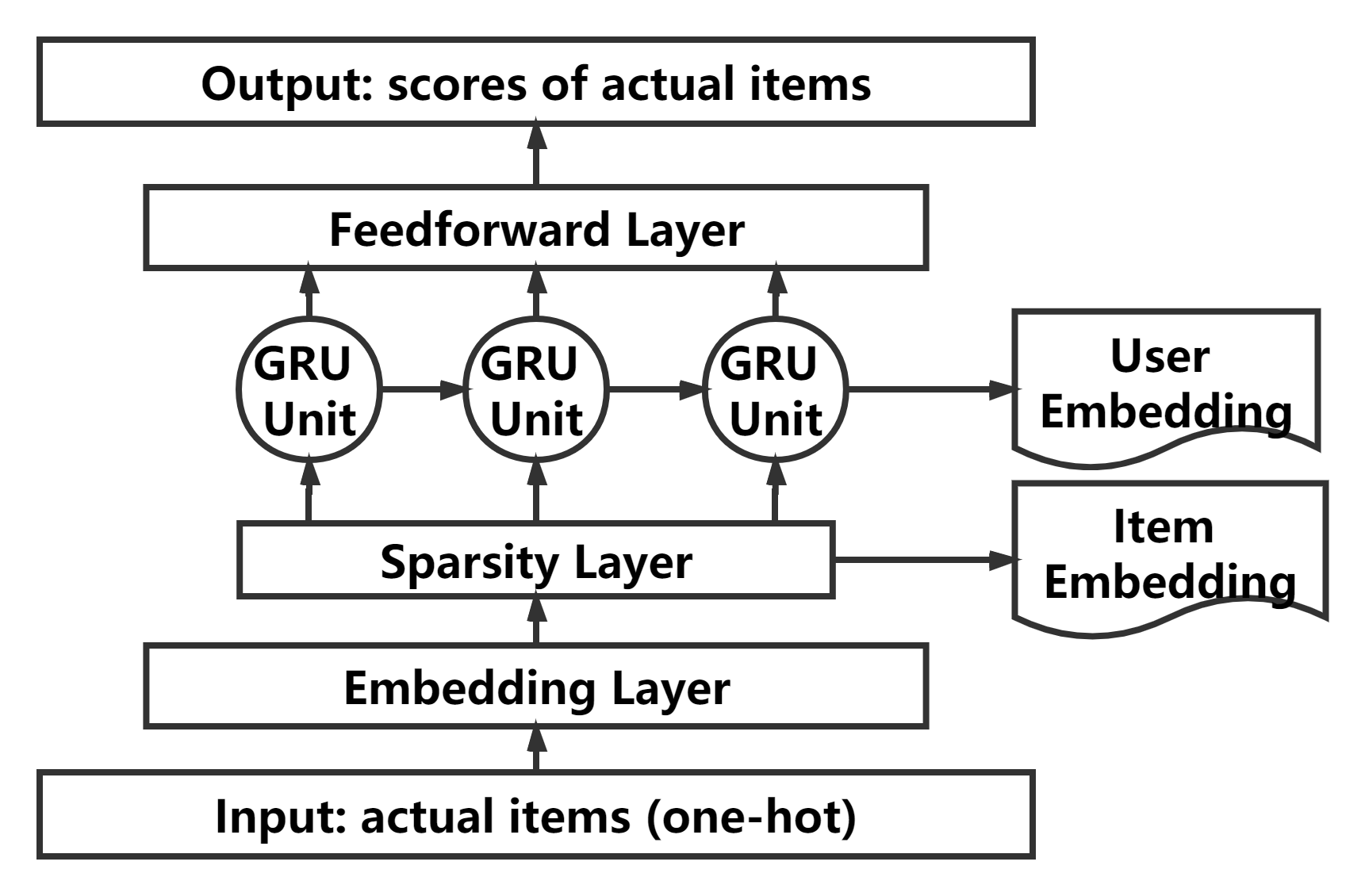}\\
        \vspace{0.02cm}
    \end{minipage}%
}%
\subfigure[]{
    \begin{minipage}[t]{0.45\textwidth}
        \centering
        \includegraphics[width=\textwidth]{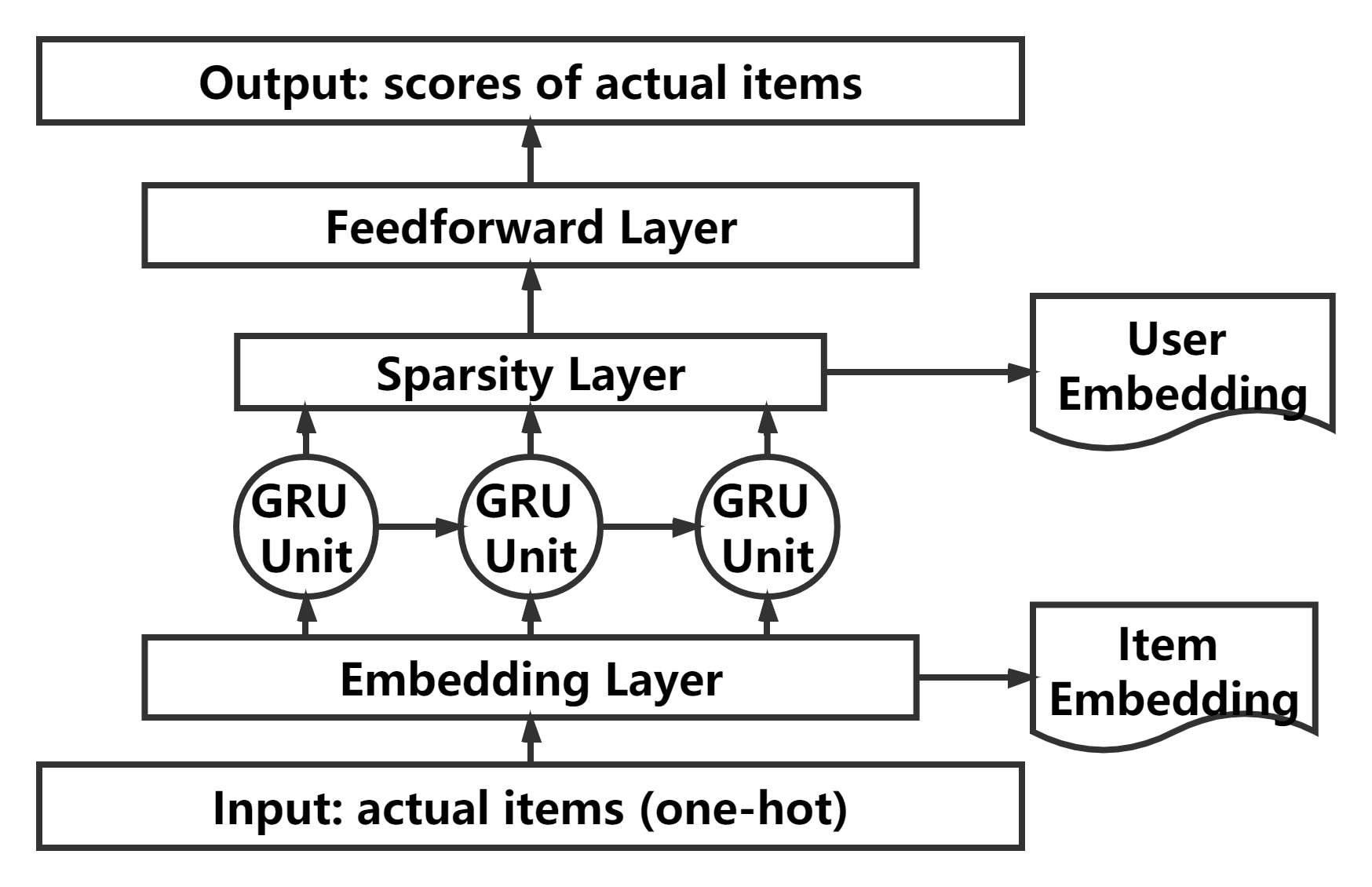}\\
        \vspace{0.02cm}
    \end{minipage}%
}%
\caption{Recommendation Model}
\vspace{-0.2cm}
\label{GRU}
\end{figure}

Inspired by GRU4REC \cite{hidasi2015session}, we use a Gated Recurrent Unit (GRU) based deep neural network to model the temporal information behind the behavior histories of users and then extract user embeddings. The reason why we use GRU instead of classic RNN units or LSTM\cite{hochreiter1997long} is that GRU based models perform better than other units on this temporal recommendation task\cite{hidasi2015session, okura2017embedding}. The detailed network framework is shown in detail in Figure \ref{GRU}.

The input of the recommendation model is a sequence of actual items in a one-hot encoding style, which is projected into a low-dimensional dense embedding through the embedding layer. In pull mode, we need to pass the dense embedding into a sparsity layer to ensure that the output item embedding is as sparse as possible, in order to reduce the communication overhead between the cloud and the client. Meanwhile, in push mode, we simply output the raw dense item embedding, for it need not to be downloaded by the client. After that, the sequence of item embeddings is passed into one or several GRU layers to mine the temporal information behind the user behavior history. If multiple GRU layers are used, the input of the next layer is the hidden states of the last layer. The output of GRU layer(s) represents the probability distribution of the next item the user will click, which is named as the user embedding. In push mode, we need to pass the user embedding into a sparsity layer to ensure that the output user embedding is as sparse as possible, in order to reduce the communication overhead between the cloud and the client. Meanwhile, in pull mode, we simply obtain the raw dense user embedding, for it need not to be uploaded to the cloud. Finally, the user embedding is passed into one or multiple feed-forward layer(s) to output the scores of actual items that the user is about to click next.

\subsection{Sparsity of Embeddings and the Recommendation Model}

There are two challenges we are supposed to tackle.  To reduce network communication overhead between the cloud and devices, we perform the Lasso regression, or $L_1$ regularization, to sparsify the item embeddings of item candidate set in pull mode or user embeddings in push mode. To reduce computational overhead on the devices, we perform an Automated Gradual Pruner (AGP)\cite{zhu2017prune} to compress the recommendation model.

We perform Lasso regression to extract the main information in item embeddings or user embeddings by sparse encoding. We formulate Lasso regression as follows.
\begin{equation}
\begin{aligned}
\hat{E}_i=\left\{
            \begin{array}{lr}
            E_i, if \left|E_i\right|>\gamma\\
            0, otherwise
            \end{array}  
            \right., \\
\mathcal{L}_{lasso}=\sum_i\left|\hat{E}_i\right|, \\
\hat{\mathcal{L}}=\mathcal{L}+\lambda_{lasso}\mathcal{L}_{lasso},
\end{aligned}
\end{equation}
where $E_i$ is the raw item embedding or user embedding, $\hat{E}_i$ is the sparse embedding, $\mathcal{L}$ is the original loss function, which is a cross entropy loss in our work, $\lambda_{lasso}$ is a weight to measure the importance of lasso penalty, and $\hat{\mathcal{L}}$ is the final loss function with lasso penalty. In the sparse layer of both pull mode and push mode of $C^3DRec$, the input embedding is truncated with a threshold $\gamma$, and only the dimensions above $\gamma$ are kept in the output embedding. Then, the lasso penalty term $\lambda_{lasso}\mathcal{L}_{lasso}$ is added to the final loss function, inducing the output embedding to be sparse.

We perform Automated Gradual Pruner (AGP)\cite{zhu2017prune} to reduce the computational overhead of training personalized models on the devices by pruning the recommendation model. Pruning is a commonly used methodology to induce \textit{sparsity} (i.e. a measure of how many elements in a tensor are exact zeros, relative to the tensor size.) in weights in deep neural networks, which assigns weights satisfying a certain criteria to zero. Note that those \textit{trimmed} weights will be "shutdown" permanently and not participate in back-propagation. A level pruner is a stable pruner where a neural network is pruned into a specific sparsity level, through a threshold magnitude criteria. AGP is an algorithm to schedule a level pruner that can be formulated as follows.
\begin{equation}
s_t=s_f+\left(s_i-s_f\right)\left(1-\frac{t-t_0}{n\Delta t}\right)^3\ for\ t\in\left\{t_0,t_0+\Delta t,...,t_0+n\Delta t\right\},
\end{equation}
where $s_t$ is the sparsity level in $t$ time step (i.e. epoch), $s_i$ is the initial sparsity level and $s_f$ is the final target sparsity level. The main idea of AGP is to initially prune the weights rapidly when redundant connections are abundant and gradually reduce the number of pruned weights when redundant connections are becoming fewer. The reason why we choose AGP is that it does not make strong assumptions on the structure of the network or layers, and it is straight-forward and not difficult to tune hyper-parameters.

\section{Evaluation}
In this section, we evaluate $C^3DRec$ from three folds: the accuracy improvement gained by personalization, the network communication overhead between the cloud and the client, and the computational overhead of on-device training.

\subsection{Dataset}
To evaluate the performance of $C^3DRec$ and baselines, we use a public user behavior dataset from Taobao~\cite{zhu2018learning}, named UserBehavior\footnote{https://tianchi.aliyun.com/dataset/dataDetail?dataId=649}, which contains nearly a million users and nearly 100 million behaviors including click, purchase, adding items to shopping cart and item favoring during November 25 to December 03, 2017. In specific, a user-item interaction contains domains including user ID, item ID, item's category ID, behavior type, and timestamp. We manually partition the interaction data into sessions by using a 706-second idle threshold. Like GRU4Rec\cite{hidasi2015session}, we filter out sessions of length 1, filter out sessions with less than 12 clicks, and filter out clicks from the test set where the item clicked is not in the train set.

To simulate the realistic scene of GDPR, we manually set a GDPR deadline $T_{device}=1512057600$ (i.e. 00:00 on December 01, 2017), after which the click data can only be used to train a personalized on one's device, rather than training the global model on the cloud. To evaluate the performance of $C^3DRec$, we manually set a training deadline $T_{test}=1512230400$ (i.e. 00:00 on December 03, 2017), after which the click data can only be used to test, rather than training whether the global model or a personalized model.

To evaluate the generalization of $C^3DRec$ on new users, we manually partition users into old users and new users by the average of timestamps of a user's clicks, i.e. users with a quantile of his/her timestamps mean above 0.9 is assigned to new users while others are assigned to old users. New users appear themselves after the GDPR deadline $T_{device}$, which means that a new user' click can only be used to train his personalized model rather than the global model. Note that old users' click are used to train both the global model and their own personalized models. After the above filtering and partitions, the scale and statistics of UserBehavior are shown in Table \ref{dataset}.
\begin{table}
  \caption{Description of the dataset}
  \label{dataset}
  \begin{tabular}{ccccc}
    \toprule
    Scale&Users&Items&Clicks\\
    \midrule
    Global training (old users) & 876,914 & 922,390 & 51,908,146\\
    Personalized training and testing(old users) & 788,472 & 886,658 & 17,200,197\\
    Personalized training and testing(new users) & 97,199 & 567,213 & 2,810,257\\
  \bottomrule
\end{tabular}
\end{table}

\subsection{Metrics and Implementation}
Like GRU4Rec\cite{hidasi2015session}, we choose Recall@20 and MRR@20 (Mean Reciprocal Rank) for evaluation. Recall@20 is the proportion of instances with the ground truth next-click item among the top-20 predicted items in all instances and does not take the ranks of the ground truth next-click items into consideration. MRR@20 is the average of reciprocal ranks of ground truth next-click items and takes the rank of their ranks into consideration.

We implement the $C^3DRec$ prototype based on the popular TensorFlow\cite{abadi2016tensorflow} to train a global model on the cloud and TensorFlow.js\cite{smilkov2019tensorflow} to train a personalized model on the device and deploy it with the WebView of an experimental application\cite{ma2019moving}. For the global training on the cloud, we take advantage of TensorFlow, and hyper-parameters are chosen carefully by running experiments at several combinations of random points of the hyper-parameter space. We choose the optimal hyper-parameter combination by selecting the highest Recall@20 and MRR@20 on the validation set (which is manually divided from the global training set). The number of GRU layers is one, with 100 hidden units in it. The batch size is 50 and the loss function is cross-entropy. The embedding size of the embedding layer is 10000. The optimization method is Adagrad\cite{duchi2011adaptive}. The learning rate is 0.01, both momentum and weight decay are 0.

For the personalized training on the devices, we take advantage of TensorFlow.js and WebView. TensorFlow.js is a deep learning framework to build and execute deep learning based models in JavaScript to run in a web browser or the Node.js environment. WebView is a control for displaying web content inside Android applications based on the Chrome engine, which provides a built-in JavaScript parser. We leverage WebView to implement the adaptation of TensorFlow.js interpreter inside Android applications, including bidirectional calling of both the Web environment and the Android application, and dynamically loading Android files in the Web environment. Besides, instead of packaging the model file (.pth) together with the application installation package (.apk), we dynamically loads the individual compressed model file as a configuration, so as to keep the installation package light-weight and flexible. In specific, we store the model file and other configuration files in the external storage directory of Android. When the application starts, our framework reads the configuration file from the storage and registers the corresponding model metadata. When the on-device training starts, our framework decompresses and loads the global model according to the file path in model metadata. We use the Google Pixel 2 smartphone model (Android 8.0, Octa-core (4x2.45 GHz + 4x1.9 GHz) Kryo, Adreno 540) to conduct our experiments on the cost of on-device training.

\subsection{Accuracy of Personalized Model}
\begin{figure}[htbp]
  \centering
  \includegraphics[width=0.6\textwidth]{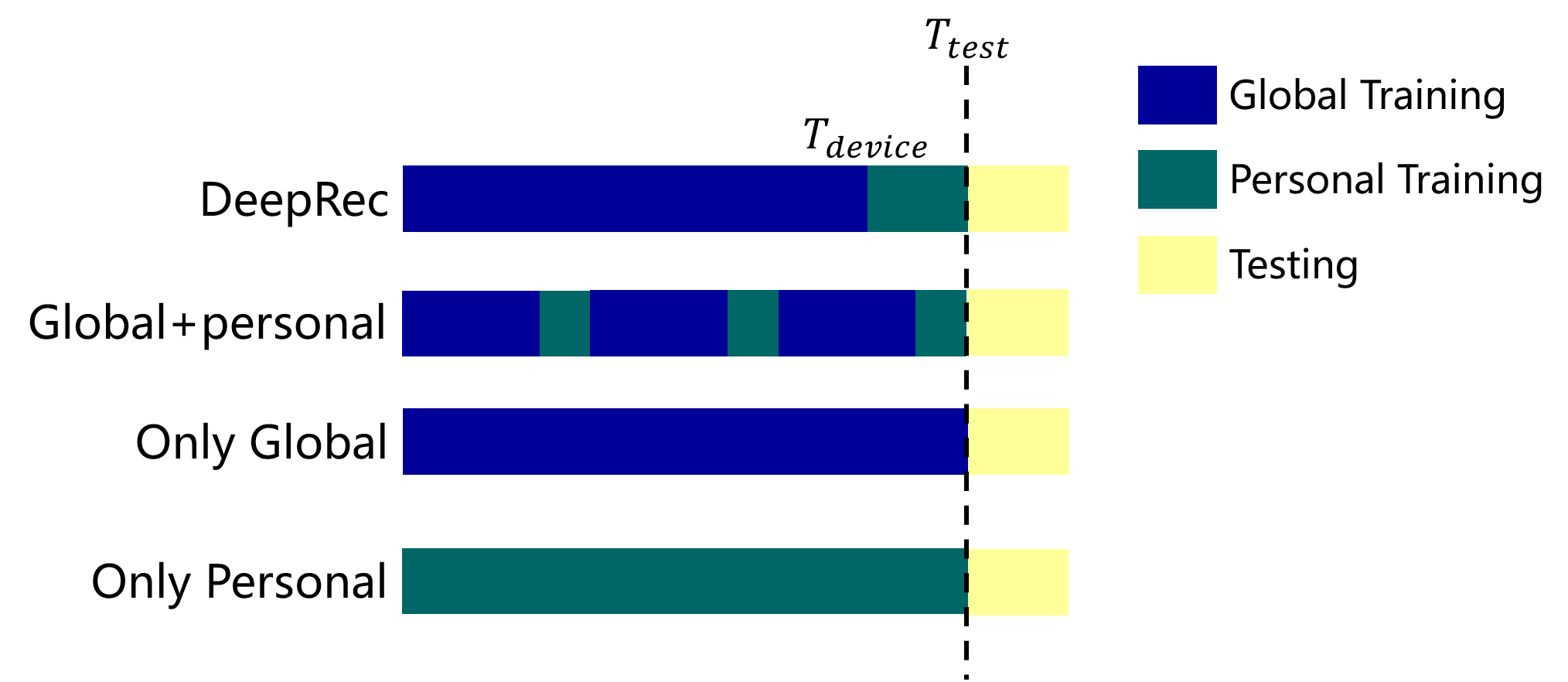}
  \caption{An illustration of the training and testing timeline}
  \label{timeline}
  \Description{}
\end{figure}

\begin{figure}[htbp]
\centering
\subfigure[Recall@20]{
    \begin{minipage}[t]{0.45\textwidth}
        \centering
        \includegraphics[width=\textwidth]{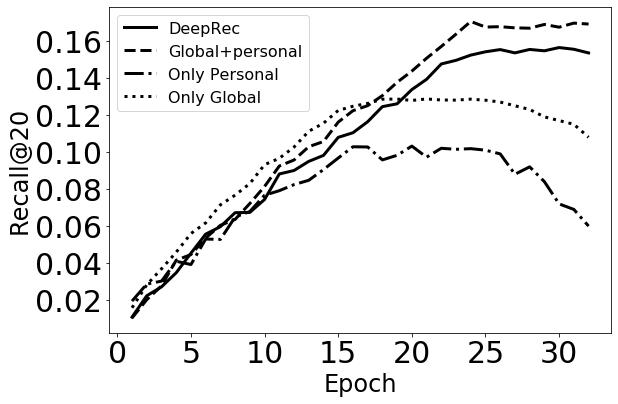}\\
        \vspace{0.02cm}
    \end{minipage}%
}%
\subfigure[MRR@20]{
    \begin{minipage}[t]{0.45\textwidth}
        \centering
        \includegraphics[width=\textwidth]{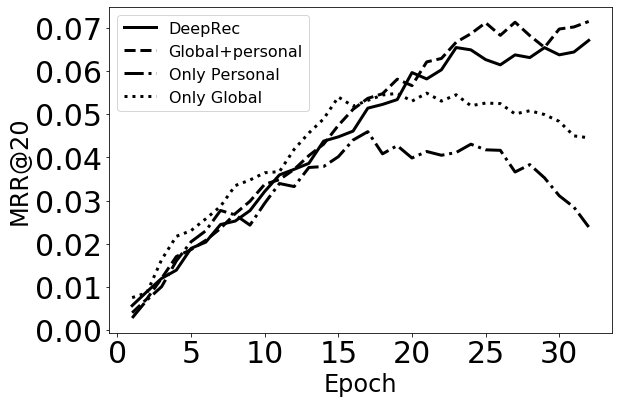}\\
        \vspace{0.02cm}
    \end{minipage}%
}%
\caption{The learning curve of $C^3DRec$ and baselines}
\vspace{-0.2cm}
\label{learning curve}
\end{figure}

\begin{table}
  \caption{Accuracy of our approach and baselines}
  \label{recall and mrr}
  \begin{tabular}{ccc}
    \toprule
    Approach&Recall@20&MRR@20\\
    \midrule
    Global + Personal & 0.168 & 0.069\\
    $C^3DRec$ (Our Approach) & 0.155 & 0.064\\
    Only global (GRU4Rec) & 0.128 & 0.052\\
    Only personal & 0.101 & 0.043\\
  \bottomrule
\end{tabular}
\end{table}
We evaluate the effectiveness of both the global models and personalized models. Figure \ref{timeline} illustrates the training and testing schedule of $C^3DRec$ and three other baselines. In this figure, \textit{$C^3DRec$} indicates to train a global model with all users' clicks in the first 7 days and fine-tune a personalized model for each user with his/her click on the 8th day, which is suitable for recommendation in the post-GDPR era where GDPR takes effect in the $8_th$ day. \textit{Global + personal} models indicate to both train a global model and fine-tune personalized models for all users with all users' clicks in the first 8 days, which is not suitable for GDPR because GDPR prohibits uploading click data on the $8_th$ day without users' consent, however, this is the upper bound that frameworks with both global models and personal models can achieve. \textit{Only personal} models indicate to only train a personalized model for each user with his/her click in the first 8 days. The \textit{only global} model indicates to only train a global model for all users with all users' clicks in the first 8 days, which is the training strategy for almost all state-of-the-art approaches for temporal recommendation, as far as we are concerned. In Figure \ref{learning curve}, we show the learning curves of Recall@20 and MRR@20 on testing set. We notice that \textit{only personal} models and \textit{only global} models converges earlier than mixture models. However, \textit{only personal} models suffer from a very poor accuracy, because a single user's training data is so few that personal models are easy to over-fit and converge to a local optimal point. \textit{Our approach} performs much better than the \textit{only global} model, i.e. GRU4Rec\cite{hidasi2015session}, because our approach both takes advantage of common click patterns through the global model and mines a user's unique click pattern through his/her personal model. Although \textit{global + personal} models actually predicts a little better than our approach, they are not suitable for the post-GDPR era, as mentioned above. The detailed recommendation accuracy are shown in table \ref{recall and mrr}.

\begin{figure}[htbp]
\centering
\subfigure[Recall@20]{
    \begin{minipage}[t]{0.45\textwidth}
        \centering
        \includegraphics[width=\textwidth]{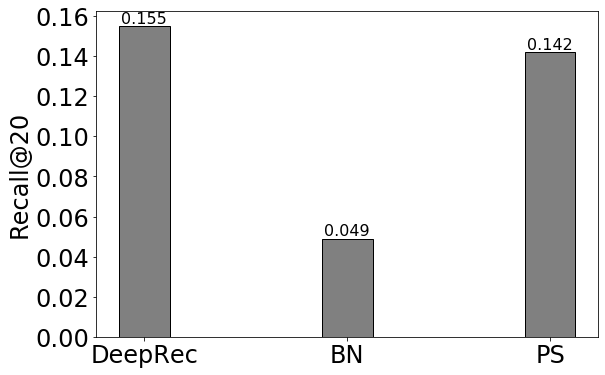}\\
        \vspace{0.02cm}
    \end{minipage}%
}%
\subfigure[MRR@20]{
    \begin{minipage}[t]{0.45\textwidth}
        \centering
        \includegraphics[width=\textwidth]{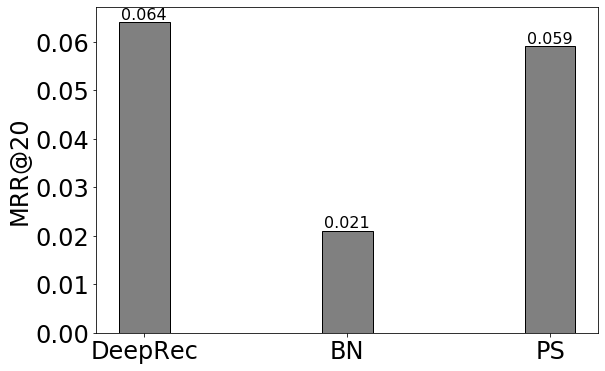}\\
        \vspace{0.02cm}
    \end{minipage}%
}%
\caption{The necessity of click history data}
\vspace{-0.2cm}
\label{login}
\end{figure}

To answer the question that if only business necessary transactional data, i.e. purchase behaviors, can predict the item a user click next, we train our model with only transactional data and only click data, to evaluate the effectiveness of click data. As shown in Figure \ref{login}, the accuracy of the model trained with only click data is close to that of $C^3DRec$, however, the accuracy of the model trained with only transactional data is much lower than that of $C^3DRec$. The explanation is rather straight-forward, the task of temporal recommendation is to provide the next click item based on the history of user behaviors, therefore, it makes no sense to predict the probability of \textit{click} based on \textit{purchase} data. Therefore, we cannot use only business necessary transactional data, i.e. purchase behaviors, to achieve an accurate temporal recommendation.

\begin{figure}[htbp]
\centering
\subfigure[Recall@20]{
    \begin{minipage}[t]{0.45\textwidth}
        \centering
        \includegraphics[width=\textwidth]{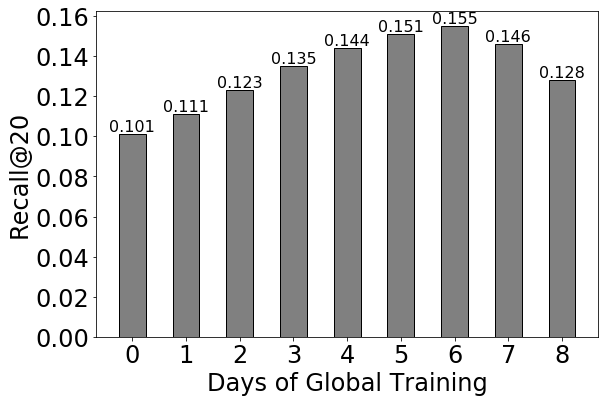}\\
        \vspace{0.02cm}
    \end{minipage}%
}%
\subfigure[MRR@20]{
    \begin{minipage}[t]{0.45\textwidth}
        \centering
        \includegraphics[width=\textwidth]{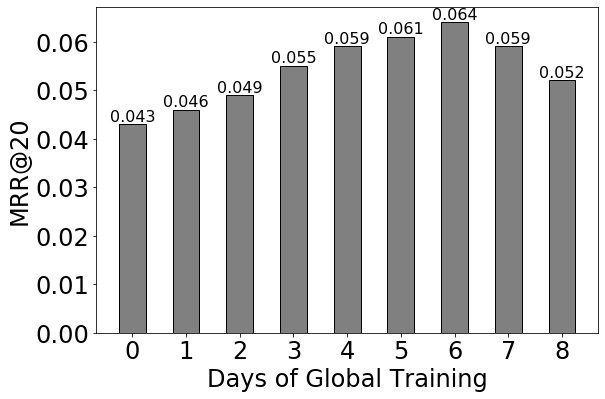}\\
        \vspace{0.02cm}
    \end{minipage}%
}%
\caption{The influence of the size of global training set}
\vspace{-0.2cm}
\label{pre-training}
\end{figure}

To measure how GDPR deadline $T_{device}$, i.e. the size of global training set and personalized training set, affects the recommendation accuracy, we scale both the time span of global training set and personalized training set by ranging $T_{device}$ from 0 days to 8 days. The detailed recommendation accuracy is shown in Figure \ref{pre-training}, where 0-day global training is actually \textit{only personalized} models, and 8-day global training is actually the \textit{only global} model, as mentioned above. The results indicates that GDPR deadline $T_{device}$ would affect the recommendation accuracy to some extent, the optimal $T_{device}$ comes at some particular point between \textit{only global} model and \textit{only personalized} models.

\begin{figure}[htbp]
\centering
\subfigure[Recall@20]{
    \begin{minipage}[t]{0.45\textwidth}
        \centering
        \includegraphics[width=\textwidth]{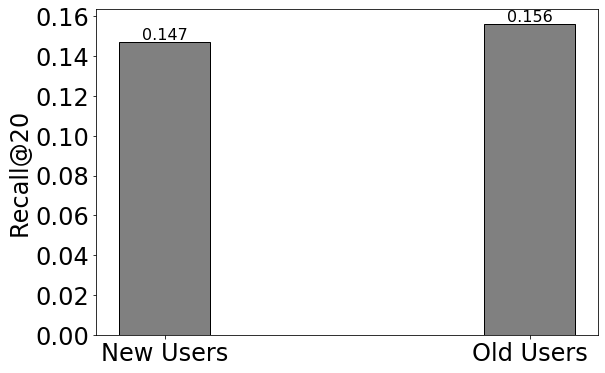}\\
        \vspace{0.02cm}
    \end{minipage}%
}%
\subfigure[MRR@20]{
    \begin{minipage}[t]{0.45\textwidth}
        \centering
        \includegraphics[width=\textwidth]{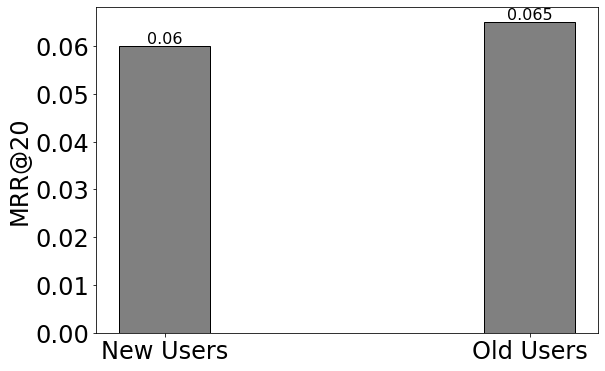}\\
        \vspace{0.02cm}
    \end{minipage}%
}%
\caption{The generalization of $C^3DRec$ on new users}
\vspace{-0.2cm}
\label{new_users}
\end{figure}

To measure the generalization of $C^3DRec$ on new users, we measure the recommendation accuracy of old users and new users, as shown in Figure \ref{new_users}. The results indicates that although new users' click history is not used for global training, and only their click history after $T_{device}$ is used for personalized training, $C^3DRec$ is able to transfer the click pattern of old users in global training to the click pattern of new users in personalized training and testing, and the global model of old users can easily generalizes to the personalized models of new users.

\begin{figure}[htbp]
\centering
\subfigure[Recall@20]{
    \begin{minipage}[t]{0.45\textwidth}
        \centering
        \includegraphics[width=\textwidth]{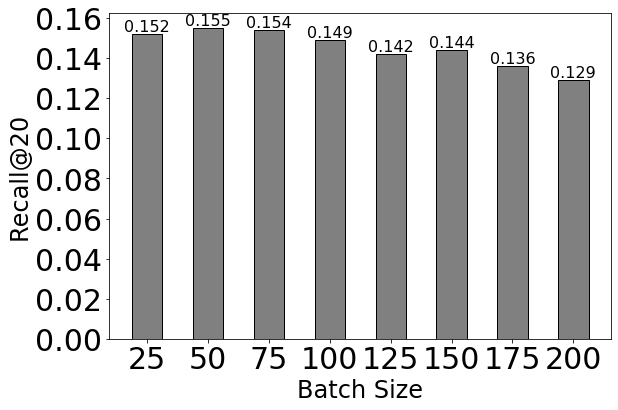}\\
        \vspace{0.02cm}
    \end{minipage}%
}%
\subfigure[MRR@20]{
    \begin{minipage}[t]{0.45\textwidth}
        \centering
        \includegraphics[width=\textwidth]{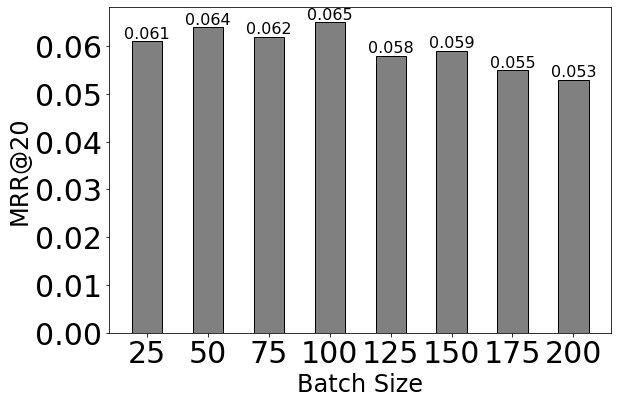}\\
        \vspace{0.02cm}
    \end{minipage}%
}%
\caption{The influence of update intervals of fine-tuning personalized models}
\vspace{-0.2cm}
\label{batch_size}
\end{figure}

Note that personalized training is not supposed to be performed once new interactions appear on the device, because users' devices are not always idle and available for training which consumes many computational resources and battery. However, if the personalized model is not updated for a long time, the recommendation accuracy is tend to drop. To measure how the update intervals influences the accuracy, we scale the size of mini-batch from 25 to 200(the personal model is not updated until a mini-batch comes to its end). As shown in Figure \ref{batch_size}, within a relatively wide range, the intervals of fine-tuning personalized models influence little on accuracy, which indicates that even if a user keep his/her personalized model on device unchanged for a relatively long period of time, it would not affect his/her recommendation user experience much.

\subsection{Communication Overhead}
There are mainly two parts of network communication overhead between the cloud and the client. In pull mode, the device needs to download the recommendation item candidate set from the cloud, while in push mode, the device needs to upload the user embedding to the cloud. In both pull mode and push mode, the device needs to download the global recommendation model from the cloud. As mentioned before, we use Lasso regression ($L_1$ regularization) to sparsify item embeddings or user embeddings, and AGP to sparsify the global recommendation model.

\begin{figure}[htbp]
\centering
\subfigure[Pull Mode]{
    \begin{minipage}[t]{0.45\textwidth}
        \centering
        \includegraphics[width=\textwidth]{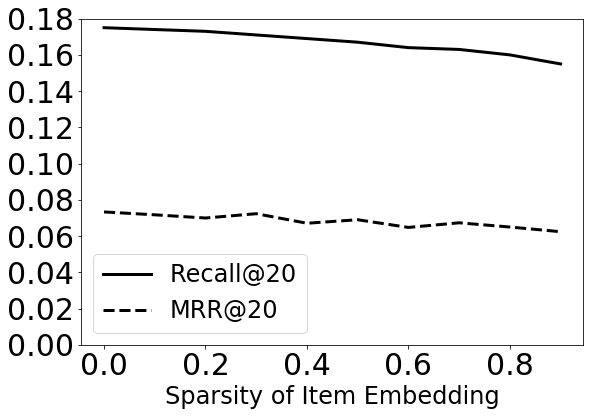}\\
        \vspace{0.02cm}
    \end{minipage}%
}%
\subfigure[Push Mode]{
    \begin{minipage}[t]{0.45\textwidth}
        \centering
        \includegraphics[width=\textwidth]{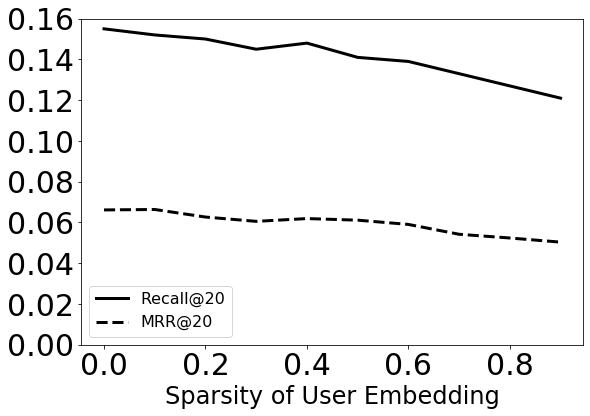}\\
        \vspace{0.02cm}
    \end{minipage}%
}%
\caption{The influence of lasso regularization}
\vspace{-0.2cm}
\label{sparsity_of_embedding}
\end{figure}

To measure how the sparsity of item embeddings in \textit{pull mode} and user embeddings in \textit{push mode} affects the recommendation accuracy, we scale the target sparsity of embeddings and compare recall@20 and MRR@20. As shown in Figure \ref{sparsity_of_embedding}, as the sparsity of item embeddings or user embeddings increases, the recommendation accuracy does not suffer from a significant drop, which encourages us to choose an aggressive embedding sparsity of 90\%, to reduce the communication overhead between the cloud and the device.

\begin{figure}[htbp]
    \begin{minipage}[t]{0.45\textwidth}
        \centering
        \includegraphics[width=\textwidth]{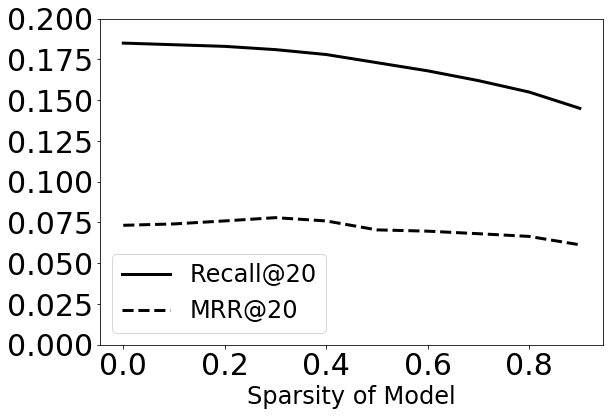}
        \caption{Sparsity of the model}
        \label{sparsity_of_model}
        \vspace{0.02cm}
    \end{minipage}%
    \begin{minipage}[t]{0.45\textwidth}
        \centering
        \includegraphics[width=\textwidth]{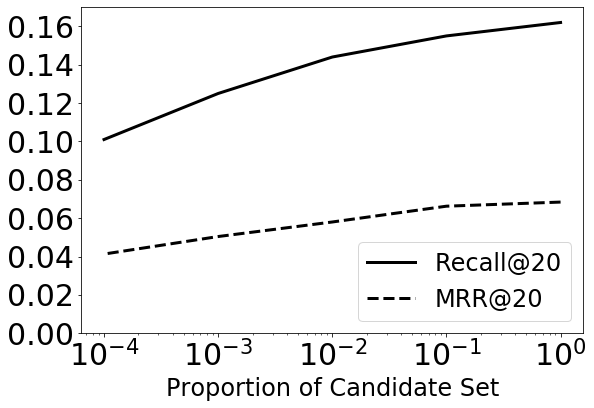}
        \caption{Size of candidate set}
        \label{proportion_of_candidate_set}
        \vspace{0.02cm}
    \end{minipage}%
\vspace{-0.2cm}
\end{figure}

To measure how the sparsity of the global recommendation model affects the recommendation accuracy, we scale the target sparsity of the global recommendation model and compare recall@20 and MRR@20. As shown in Figure \ref{sparsity_of_model}, as the sparsity of the model increases, the recommendation accuracy does not suffer from a significant drop, which encourages us to choose an aggressive model sparsity of 90\%, to reduce the communication overhead between the cloud and the device.

Note that we also use an item-CF based filtering model to reduce the computational overhead of ranking the combination of a given user embedding and recommendation item embeddings. To measure how the size of the candidate set affects the recommendation accuracy, we scale the proportion of the candidate set over all items and compare recall@20 and MRR@20. As shown in Figure \ref{proportion_of_candidate_set}, it turns out that the recommendation accuracy would drop dramatically as the proportion of candidate set becomes smaller. Here, we carefully choose a conservative proportion of 10\%, in order to make a trade-off between the computational overhead and the recommendation accuracy.

\subsection{Computational Overhead}
\begin{table}
  \caption{Cost of on-device training in pull mode}
  \label{ondevice_pull}
  \begin{tabular}{cccc}
    \toprule
    Metric&$C^3DRec$&$C^3DRec$ w/o AGP&$C^3DRec$ w/o item embedding\\
    \midrule
    Training Time&1163 seconds&3725 seconds&--\\
    CPU\%&66.3\%&65.8\%&--\\
    Memory&842.9 MB&844.1 MB&$\infty$\\
    Downloaded Item Candidate Set&391.1 MB&391.1 MB&0.4 MB\\
    Downloaded Model&1.71 MB&16.2 MB&156.8 MB\\
  \bottomrule
\end{tabular}
\end{table}
\begin{table}
  \caption{Cost of on-device training in push mode}
  \label{ondevice_push}
  \begin{tabular}{cccc}
    \toprule
    Metric&$C^3DRec$&$C^3DRec$ w/o AGP&$C^3DRec$ w/o embedding sparsity\\
    \midrule
    Training Time&1117 seconds&3706 seconds&1123 seconds\\
    CPU\%&66.9\%&65.5\%&66.2\%\\
    Memory&479.8 MB&480.9 MB&475.6 MB\\
    Uploaded User Embedding&4.2 KB&4.2 KB&40.0 KB\\
    Downloaded model&1.71 MB&16.2 MB&1.69 MB\\
  \bottomrule
\end{tabular}
\end{table}
We use several techniques to make on-device training more feasible and less resource-consuming, including AGP, item embedding, and embedding sparsity. To evaluate how each technique affects the feasibility and computational overhead of on-device training, we conduct ablation experiments both in pull mode and push mode. Note that \textit{$C^3DRec$ w/o item embedding} means removing the embedding layer and use one-hot encoding to represents items. As shown in Table \ref{ondevice_pull}, $C^3DRec$ without item embedding cannot be deployed on devices because the memory it needs to deploy and train is dramatically beyond that of mainstream mobile devices. In addition, AGP helps to both save training time and reduce the size of the downloaded global model. As shown in Table \ref{ondevice_push}, item sparsity helps reduce the size of the uploaded user embedding and AGP has similar benefits to pull mode. However, the effect of embedding sparsity in push mode is not as significant as that of item embedding in pull mode, because the device does not need to download the item candidate set from the cloud, which dramatically reduces the communication overhead. However, the push mode has its own problems that it still suffers from potential risks of partial leakage of user privacy, because it requires to upload the intermediate results of the recommendation model (i.e. the user embedding). Still, the privacy leakage risk of push mode is much lower than that of current approaches that upload all user behavior histories to the cloud, therefore, users are more likely to agree with privacy consents in push mode, compared with current frameworks. We conclude the advantages and disadvantages of pull mode and push mode in Table \ref{comparison}.
\begin{table}
  \caption{Comparison of pull mode and push mode}
  \label{comparison}
  \begin{tabular}{p{0.12\columnwidth}p{0.18\columnwidth}p{0.3\columnwidth}p{0.3\columnwidth}}
    \toprule
    &Current Approaches&$C^3DRec$ in pull mode&$C^3DRec$ in push mode\\
    \midrule
    Require a Global Model&Yes&Yes&Yes\\
    Require Personal Models&No&Yes&Yes\\
    Communication Overhead&No&Download the Global Model and Download the Item Candidate Set (Large)&Download the Global Model and Upload the User Embedding (Small)\\
    Computational Overhead&No&Training Personal Models&Training Personal Models\\
    Privacy Leakage&Require to Upload Raw Data (High Risk)&No&Require to UpLoad the User Embedding (Low Risk)\\
  \bottomrule
\end{tabular}
\end{table}

\section{Discussion}
The accuracy improvement of $C^3DRec$ reveals an interesting finding that \textbf{after training a global model with all users' data, fine-tuning a personalized model for each user can significantly improve the recommendation accuracy.} However, due to the time limit, we could not explore the underlying reasons for this accuracy improvement. We assume that different kinds of users might have different behavior patterns and personalized models are able to model those behavior patterns case by case, therefore, this framework predicts the next click item with a higher accuracy. We plan to mine the underlying reasons for this accuracy improvement and detect the concept drift\cite{gama2014survey} when users' behavior patterns have a sudden change. When a concept drift is detected, we incrementally shift from the current recommendation model to another model with a higher accuracy on a certain window of training data, and this is an online learning\cite{al2018dynamic} or incremental learning\cite{yang2019adaptive} manner. Essentially, incremental learning is to train a personalized model for each stable stages between two concept drifts, instead of for each user (as in $C^3DRec$). Therefore, compared with $C^3DRec$, incremental learning based recommendations could significantly reduce the number of personalized models and therefore reduce the overall computational overhead, while keeping the recommendation accuracy from dropping much (but still higher than the \textit{only global} model). However, incremental learning based recommendation is not suitable for the post-GDPR era, therefore, it is not the scope of this paper, and we will explore that in our future work.

Another problem lies in the sustainability of $C^3DRec$. As shown in Figure \ref{pre-training}, before the $7^th$ day, the recommendation accuracy increases as the size of the global training set increases, which indicates that $C^3DRec$ still requires a considerable amount of behavior data to train the global model. In an extreme situation when personal models are trained for a much longer period than the global model, our $C^3DRec$ will degenerate to \textit{only personal} model, which suffers from an unstable accuracy. This problem can be temporarily relieved by encouraging \textit{some} users agree to upload their behavior data in the consent by certain incentives. We can update the global model with the new behavior data of those users and download it to \textit{other} users' devices for personal training. To solve this problem, we have to explore the emerging federated learning (FL)\cite{konevcny2016federated, bonawitz2019towards, yang2019federated} and deploy FL in the recommendation settings in our future work.

\section{Related Work}
In 1992, Goldberg et al. proposed a user based collaborative filtering algorithm \cite{goldberg1992using}, whose main idea is that users with similar historical behaviors are more likely to purchase the same products. In 2001, Sarwar et al. proposed an item-based collaborative filtering algorithm \cite{sarwar2001item}, whose main idea is that a user is likely to purchase an item that shares similar features, i.e. user-item interactions, with items he purchases before. The advantages of collaborative filtering models are that they do not require to mine the content features of items, they can effectively utilize the feedback information of similar users, and they can discover the potential preferences of users by recommending new items that have not appeared in their history records before. However, collaborative filtering suffers from cold start, sparsity and scalability problems.

With the emerge of deep learning techniques, researchers focus on applying deep learning techniques to recommendation systems to improve its performance. Nonetheless, most deep learning based approaches are still variants and/or hybrids of collaborative filtering. In 2016, Cheng et al. from Google proposed the Wide and Deep Learning algorithm for recommending systems \cite{cheng2016wide}. The \textit{wide} model takes one-hot encoded binary features as input, and mines the correlation between items or features from the historical information, whose recommendation result is often the item directly related to the items in the historical record. The \textit{deep} model takes numerical features and dense embeddings of continuous features as input, which learns new feature combinations and is able to recommend an item that has never existed in the historical record. Jointly training the Wide model and Deep model allows the framework both to memorize and to generalize. However, in order to mine cross-features, the \textit{wide} model requires an explicit feature engineering. In order to learn the low-order and high-order cross-features in an end-to-end manner, in 2017, Guo et al. proposed the DeepFM model \cite{guo2017deepfm}. DeepFM uses a Factorization Machine (FM) model to take the place of the \textit{wide} model, which shares the high-dimensional sparse features and low-dimensional dense features as input with the \textit{deep} model. The FM extracts second-order cross-features and the \textit{deep} model extracts high-order cross-features automatically.

The user data collected by mainstream business recommendation platforms usually contains timestamps, however, both CF-based traditional approaches and CF-based deep learning models cannot model the temporal information behind the sequential dataset, because CF-based frameworks treat each item independently of the other items appeared in the same period, which cannot model a user's continuous preference trends through time. To solve this problem, Hidasi et al. proposed the GRU4REC\cite{hidasi2015session} to leverage recurrent neural networks (RNN) to extract the temporal information behind a user's preferences over items. GRU4Rec takes the click sequence in a period, i.e. a session, makes several adaptions to allow RNN based model fit for the recommendation setting, trains the model with point-wise or pair-wise loss\cite{rendle2012bpr}, and finally outputs the possibility of each item to be clicked next. In recent years, many researchers focus on improving the performance of GRU4Rec. For example, Hidasi et al. takes unstructured data such as images and texts as features\cite{hidasi2016parallel}. Bogina et al. takes the user's dwell time on items into consideration\cite{bogina2017incorporating}. Quadrana et al. takes the user's identity into consideration and construct a hierarchical RNN with the user-level GRU and the session-level GRU\cite{quadrana2017personalizing}. Jannach et al. ensembles the recommendation results of both an RNN based model and a K-nearest neighbor (KNN) based model\cite{jannach2017recurrent}.

However, the above approaches all require to train a global model by collecting all users' to the cloud, which would be impractical without the user's consent in the post-GDPR era. DeepType \cite{xu2018deeptype, xu2018deepcache, xu2019deepwear} made an effort to solve this dilemma between accuracy and privacy concerns by training a global model on the cloud using massive public corpora, and then incrementally customizing the global model with data on individual devices. However, this framework can not be directly deployed to recommendation settings, due to the heterogeneity of item metadata. There is hardly any public dataset of cross-platform recommendation, to extract common prior knowledge for most recommendation systems. Besides, CCMF \cite{gao2019privacy} preserves user privacy operated by different companies with a differential privacy based protection mechanism and utilizes a CF-based model to perform the cross-domain location recommendation. However, this model cannot be directly applied to deep learning based models because they could suffer more from computational overhead and time consumption with a differential privacy mechanism, and CF-based models could not fully model the temporal information in our setting. An intermittent learning framework\cite{lee2019intermittent} is proposed to execute certain types of machine learning tasks effectively and efficiently with a carefully selected and optimized training and inference process. However, because it requires to train machine learning based models from scratch and perform few specialized optimizations on structures of deep neural networks (like tensors, filters, RNN cells), intermittent learning is limited to classic machine learning models and small-scaled neural networks.

\section{Conclusion}
We propose $C^3DRec$, a cloud-client cooperative deep learning framework of mining interaction behaviors for recommendation while preserving the user privacy. $C^3DRec$ supports pull mode, where the item candidate set is pulled from the cloud, and push mode, where the user embedding is uploaded to the cloud. To reduce both communication and computational overhead of devices, we introduce an Automated Gradual Pruner and a Lasso regression in our $C^3DRec$ framework to induce sparsity. We evaluate the performance of $C^3DRec$ on user behaviors from a large-scale public dataset, and the experimental results show that we achieve up to 10x reduction in communication overhead and reduce the computational overhead towards the range of middle-class mobile devices, with minimal loss in accuracy. In our future work, to mine and take advantage of the underlying reasons for the accuracy increase of fine-tuning personalized models, we will explore incremental learning based recommendations. In addition, to maintain the recommendation sustainability of $C^3DRec$, we will explore federated learning based recommendations.

\bibliographystyle{ACM-Reference-Format}
\bibliography{ref}

\end{document}